\documentclass[rnote]{aa} 

\usepackage{graphicx}
\usepackage{natbib}
\usepackage{amsmath}
\usepackage{color}

\usepackage{txfonts}
\begin{document}

   \title{A fast version of the 
k-means classification algorithm for astronomical applications}


   \author{I.~Ordov\'as-Pascual\inst{1,2}
          \and
          J.~S\'anchez~Almeida\inst{3,2}  }

\authorrunning{Ordov\'as-Pascual \& S\'anchez~Almedia}

   \institute{
Instituto de Física de Cantabria, Avenida de los Castros, s/n
E-39005 Santander, Spain\\
              \email{ordovas@ifca.unican.es}
         \and
              Departamento de Astrof\'\i sica, Universidad de La Laguna, Tenerife, Spain
         \and
              Instituto de Astrof\'\i sica de Canarias, E-38205 La Laguna,Tenerife, Spain\\
             \email{jos@iac.es}
             }

   \date{Received March 14, 2014; accepted: March 28, 2014}

 
  \abstract
   {K-means is a clustering algorithm that has been used 
   to classify large datasets in astronomical 
     databases. It is an unsupervised method, 
     able to cope very different types of problems.}
   {We check whether a variant of the algorithm called single pass 
   k-means can be used as a fast alternative to the traditional k-means. 
}
{
The execution time of the two algorithms are compared when classifying subsets 
drawn from the SDSS-DR7 catalog of galaxy spectra.
}
   {Single-pass k-means turn out to be between 
20\,\%\ and 40\,\% faster than k-means
and provide statistically equivalent classifications.
This conclusion can be scaled up to other larger databases because the 
execution time of both algorithms increases linearly with the number of objects.
}
{
Single-pass k-means can be safely used as a fast alternative to k-means.
}


\maketitle
   
%

\section{Rationale}\label{intro}

The volume of many existing and forthcoming astronomical databases 
is simply too large to use traditional techniques of analysis. 
Objects cannot be inspected 
individually by astronomers, and decisions about 
whether downloading 
observations from a satellite or about following 
up interesting targets will 
be taken by numerical algorithms. 
Two examples of observations that must be handled using
automatic methods are the datasets gathered by the satellite
 Gaia\footnote{http://sci.esa.int/gaia/} 
 (\citeauthor{2012AN....333..453P}~\citeyear{2012AN....333..453P})
and the images to be provided by
the Large Synoptic Survey Telescope\footnote{http://www.lsst.org/lsst/}
(LSST, \citeauthor{2008arXiv0805.2366I}~\citeyear{2008arXiv0805.2366I}).
Gaia can only download a minuscule fraction of the observed
frames, and onboard software decides what is sent back
to earth. LSST will image the full southern sky every few days,
requiring that more than 30 terabytes are processed and stored
every day during ten years.
Thus new automated techniques of analysis must be developed. 
Regardless of the 
details, the methods to be chosen are bound to be central to future
astronomy.

In this context,
our group has been using the algorithm k-means as an 
automated tool to classify large astronomical
data sets. It has been shown to be fast and robust in different 
contexts, for example, 
to improve the signal-to-noise ratio by stacking similar spectra 
\citep{2009ApJ...698.1497S},
to identity unusual objects in large datasets of galaxies and stars
\citep{2013ApJ...763...50S,2013RMxAC..42..111S}, 
to search for rare targets that are particularly telling from a physical
point of view \citep{2011ApJ...743...77M}, 
to select alike targets to speed up complex modeling
of spectro-polarimetric data \citep{2000ApJ...532.1215S}, 
to identify and discard noisy spectra \citep{2013RMxAC..42..111S}, 
or to classify large astronomical datasets \citep{2010ApJ...714..487S,2013ApJ...763...50S}.

Many other applications can be found in the literature, e.g., clustering analysis of stars  \citep{2012MNRAS.427.1153S}, spectroscopy of
H$\alpha$
objects in IC 1396 star-forming region \citep{1996A&A...311..145B}, 
study of formation of Ultracompact Dwarf Galaxies \citep{2012ApJ...750...91C}, 
detection of anomalous objects among periodic variable stars \citep{Rebbapragada:2009:FAP:1507556.1507573} 
and description of galaxy diversification \citep{2012A&A...545A..80F}.

So far we have used the traditional version of
k-means, which requires finding cluster centers and 
assigning the objects to them in a sequential way. 
There is another version of the algorithm called 
{\em single-pass k-means}
that does the finding of the clusters 
and the assignation simultaneously 
(e.g., Bishop 2006).
Because of this unification of two steps in only one,
single-pass k-means is expected to be faster (and so more efficient) 
than the traditional approach.

In this {\em Research Note} we compare the performance of 
the two variants of algorithms to see whether single pass k-means 
can be reliably used as a fast alternative to the traditional 
k-means for astronomy applications. Both algorithms are described in 
Sect.~\ref{algorithms}.The comparison is worked out
 in Sect.~\ref{tests}, and it is based on the SDSS-DR7 spectra database.
We use this dataset because it has been thoroughly tested 
with the original k-means \citep{2010ApJ...714..487S}.
Single pass k-means is indeed faster than the original
algorithm and provides statistically equivalent results,
as we conclude in Sect.~\ref{conclusions}.

%
%
\section{K-means and single-pass k-means}\label{algorithms}

In the context of classification algorithms, objects are points in a high-dimensional space 
with as many  dimensions as the number of parameters used to describe the objects. 
(For example, the 
dimension of the space is the number of wavelengths of
 the spectra used for testing in  Sect.~\ref{tests}.)
The catalog to be classified is a set of points in this space,
 and so the (Euclidean) distance between any pair 
of them is well defined. Points (i.e., objects) are assumed to be 
clustered around a number of cluster centers. 
The classification problem consists in (1) finding the number of clusters,
 (2) finding the cluster centers, 
and (3) assigning each object to one of these centers.
In the standard formulation, k-means begins by selecting 
a number k of objects at random from the full dataset. 
They are assumed to be the centers of the clusters, and  then each object in the 
catalog is assigned to 
the closest cluster center (i.e., that of minimum distance). Once all
 objects have been classified, the cluster center is recomputed as the centroid of 
the objects in the cluster. 
This procedure is iterated with the new cluster centers, 
and it finishes when no object is reclassified in two 
consecutive steps. The number of clusters k is arbitrarily chosen, 
but in practice, the results are insensitive 
to this selection since only a few clusters possess a significant number of members, 
so that the rest can be discarded. On exiting, the algorithm provides a number of clusters, their 
corresponding cluster centers, as well as the classification of all the original objects now assigned to 
one of the clusters.

As a result, the standard k-mean method is divided into 
two steps; the first one is the assignation step. The $i-th$ object $x_i$ 
is assigned to the cluster $k$ if the distance between $x_i$ and the  
$k-th$ cluster center  $\mu_k$ is less than the distances to all 
other cluster centers, 
\begin{equation}
|x_i-\mu_k| \leq |x_i-\mu_j|~ \forall j,
\end{equation} 
where the index $j$ labels all possible cluster centers.
The assignation is quantified in terms of the  matrix $J(i,j)$ defined as 
\begin{eqnarray}\label{matrix}
J(i,k)=1,&\nonumber \\
J(i,j)=0~& \mbox{ for } j \neq k.
\end{eqnarray}
Once the $n$ objects in the catalog have been assigned, the second step consists of 
computing new cluster centers as the centroids of all the objects in the classes, i.e.,
\begin{equation}\label{cluster}
\mu_k=\dfrac{\displaystyle\sum_{i=1}^n J(i,k) x_i}{N_k},
\end{equation}
with $N_k$ the number of objects assigned to class $k$, 
\begin{equation}\label{nk}
N_k=\displaystyle\sum_{i=1}^n J(i,k).
\end{equation}
The two steps are iterated until there are negligible reassignments 
between successiveiterations. In other words, when repeated until
 the assignation matrix $J(i,j)$ has negligible variation between two iterations.

The objective of the alternative {\em single-pass k-means} method
is to update the centroids on-the-fly 
immediately after the assignation of each data vector, without having to 
finish assignating all the vectors in the database.
This algorithm is expected to be faster because we  do not  
have to wait to update the cluster centroids until all 
data are reassigned.
As in the case of k-means, this new method begins by
choosing the initial centroids randomly in the 
database, and then assigns each data vector to the closest centroid. 
Then the loop that combines Steps 1 and 2 begins. 
Object number $i$ is assigned to the nearest cluster centroid. 
If that data element does not change its class, then the algorithm goes 
to the next element $i+1$. 
If it changes, the centroids of the initial class 
and the final class are recalculated 
immediately after the assignation. Assume that the $i-th$ object previously 
in class $k$ is now assigned to class $m$,
\begin{eqnarray}
J(i,k)^{new}=&J(i,k)^{old}-1,\nonumber \\
J(i,m)^{new}=&J(i,m)^{old}+1,
\end{eqnarray}
where the superscripts $old$ and $new$ refer to the value before and after the 
reassignment, respectively.
Then the centroids of the clusters are updated as \citet[][Sect.~9.1]{Bishop2006},
\begin{equation}\label{minus}
\mu_k^{new}= \dfrac{\mu_k^{old}(N^{new}_k +1)- x_i}{N^{new}_k}=
\mu_k^{old}+(\mu_k^{old}- x_i)/{N^{new}_k},
\end{equation}
\begin{equation}\label{plus}
\mu_m^{new}=\dfrac{\mu_m^{old}(N^{new}_m - 1)+ x_i}{N^{new}_m}=
\mu_m^{old}-(\mu_m^{old} - x_i) /N^{new}_m,
\end{equation}
which are just renderings of Eq.~(\ref{cluster}) with the new assignation of the $i-th$ object. 
After those two centroids are updated, the algorithm continues 
with the next data vector until completion of the catalog.
As in the regular k-means, the catalog is classified repeatedly 
until no further reassignment in needed.

%
%
\section{Tests}\label{tests}

We carried out two sets of tests to verify whether, on the
one hand, single-pass k-means are faster than k-means and, on the 
other hand, if the classifications resulting from both methods
are equivalents.  We explain these two tests 
and their results in the following sections.

The tests are based on the SDSS-DR7 spectroscopic galaxy catalog 
\citep[][]{2009ApJS..182..543A},  which we choose
because it has already been classified using k-means
\citep[][]{2010ApJ...714..487S}. 
The 
resulting classes are known in quite some detail  
\citep{2011MNRAS.415.2417A,
2012ApJ...756..163S}, so we do not show and discuss them here.
This selection implies that the classification space has 1637 dimensions
set by the number of wavelengths in the spectra.

The tests have been carried out in two rather modest computers: a 
laptop\footnote{Intel Core i5 CPU M520 240 GHz; 3.0 Gb RAM; Ubuntu 11.4.} 
(hereafter {\em laptop}) and a 
desktop\footnote{AMD Athlon(tm) 64$\times$2 Dual Core Processor 5600+; 
4.0 Gb RAM; Fedora 17 64 bits.} 
(hereafter {\em FORD}). 
Laptop and FORD  have RAM memory of 3 Gb and 4 Gb, respectively, so the datasets 
cannot be very large. This fact sets the number of objects used in the 
tests to a range between 1\,000 and 20\,000 galaxy spectra.

\subsection{Time per classification}\label{test1}

First of all, we measure the relative speed of k-means and single pass k-means
by classifying different  subsets of galaxy spectra from SDSS-DR7 
and comparing the time needed for completion.
We choose random subsets of the full SDSS-DR7 catalog having 
between 1\,000 and 20\,000 galaxy spectra.

For each subset and each algorithm, we repeat the classification
ten times to separate systematic differences between the algorithms 
from time differences due to the random initial conditions.
If the randomly chosen initial centroids are very similar to the final 
centroids, it takes much less time for any algorithm to converge.
The time differences are quantified in terms of the gain $G$,
\begin{equation}
G=100 \times  \dfrac{t_{km}-t_{spkm}}{t_{km}},
\end{equation}
where the symbols $t_{km}$ and $t_{spkm}$ denote the time per 
classification for k-means and single pass k-means, respectively.
Since classifications are repeated several times, we compute
the average and the dispersion of the gain.

\begin{figure}[h!]
   \centering
   \includegraphics[width=9cm]{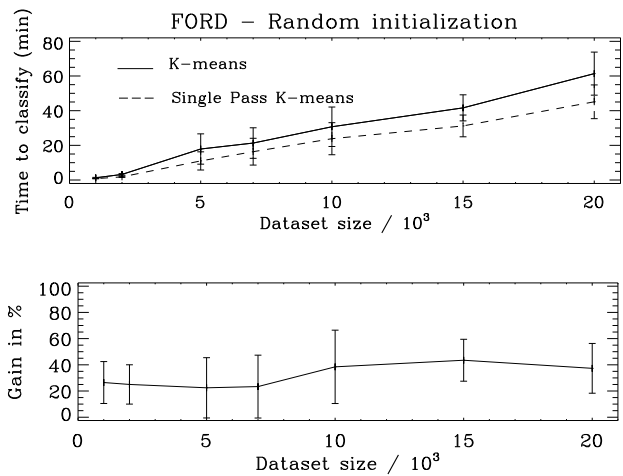}
      \caption{
Top:
Computer time required for FORD to classify 
subsets of the SDSS-DR7 galaxy spectrum catalog.
Given the number of galaxies to be classified 
(in abscissa), the time when using k-means (the solid line)
is systematically longer than the time for the alternative 
single pass k-means (the dashed line).
The computer time increases linearly with the number 
of galaxies in the catalog. Error bars code the 
dispersion produced by the random initialization of the 
algorithms.
Bottom:  
Gain when using single pass k-means, which saves
between 20\,\% and 40\,\% of the time.
              }
         \label{ford}
   \end{figure}

\begin{figure}[h!]
   \centering
   \includegraphics[width=9cm]{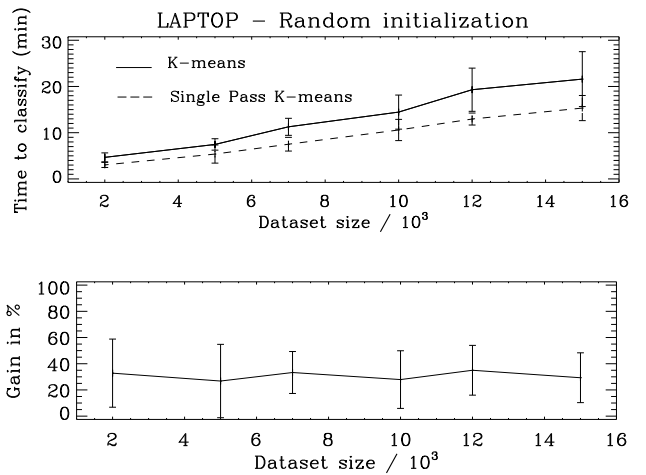}
      \caption{
Time for the classification (top) and gain (bottom) when using the laptop.
For symbols and further details, see Fig.~\ref{ford}. 
              }
         \label{alien}
   \end{figure}

The results of our test are shown in Figs.~\ref{ford}~and~\ref{alien}. 
The time for classification depends strongly on the initialization,
and this leads to a large dispersion of the time per classification. 
For example, for 20\,000 spectra  FORD's computer time varies 
from 40 and 80 minutes (Fig.~\ref{ford}). 
On top of this significant scatter, there is a systematic difference between the
two methods, where single pass k-means between 20\,\% and 40\,\% faster than k-means
-- the mean gain spans from 20\,\% to 40\,\% independently of the 
size of the dataset and the computer (see Figs.~\ref{ford}~and~\ref{alien}; 
bottom panels). This systematic gain when using single pass k-means
is the main result of our RN, provided that 
the two algorithms yield equivalent classifications. 
This equivalence is indeed
proven in Sect.~\ref{equivalence}. 

%
%
%

The tests described above required approximately seven CPU days to run. 
This limited the size of the largest subset, since the  
required time increases linearly with the number of 
objects in the catalog (Figs.~\ref{ford}~and~\ref{alien}).
However, single pass k-means would outperform k-mean even 
for other larger datasets. That the computer time employed 
by the two alternative algorithms increases linearly with time
implies that the gain should be constant even for 
significantly larger datasets. Moreover, k-means is a workhorse 
proven to converge in many very different contexts. 
The datasets we use are not special, therefore 
the properties inferred for them can be probably extrapolated to many 
other datasets.

\subsection{Equivalence of the classifications provided by the two algorithms}\label{equivalence}

K-means and single pass k-means render different classifications
of a catalog even if they start from the very same initial 
cluster centers. However, the two classifications are equivalent for practical 
applications. With a given a dataset, 
k-means does not provide a single classification but a number of them 
generated by the random initialization of the algorithm 
(see Sect.~\ref{algorithms}). This is a well known 
downside of k-means, whose impact must be evaluated
in actual applications of the algorithm 
\citep[see, e.g.,][Sect.~4]{2010ApJ...714..487S}. 
There is an intrinsic uncertainty of the classification 
ensuing from the random initialization, therefore 
the classes resulting from single pass k-means and k-means 
are equivalent so far as they are within this uncertainty. 
Consequently, to study whether single pass k-means is statistically equivalent to k-means, 
we test that the  differences between classifications carried out using 
the two methods are similar to the differences when comparing various 
initializations of the same method.

To carry out this test, we choose a subset 
of 20\,000 galaxy spectra randomly drawn from SDSS-DR7 
and then 50 different initializations. 
We proceed by classifying the 20\,000 spectra using 
those 50 initializations and the two algorithms,
so as to obtain 100 classifications of that dataset. 
It took approximately four days for FORD to complete the task.
The idea is to compare these classifications pairwise, and we  
do it by employing the parameter {\em coincidence} defined by 
\citet[][Sect.~2.1]{2010ApJ...714..487S}.  In essence, the 
classes in two classifications are 
paired so that they contain the most objects in common. The percentage of 
objects in these equivalent classes is the {\em coincidence}, which would 
be 100\,\% if the two classifications were identical.

The 100 classifications can be paired in 5050 different ways with some combining 
only k-means classifications, some combining only single pass k-means
classifications, and some mixing them up. They can be divided into four groups:
\begin{enumerate}
  \item 50 pairs of classifications where  each member of the pair 
has been treated with a different algorithm but both with the 
same initialization, \label{g1}
  \item 2450 pairs of classifications where each member has been treated with a  
different algorithm and a different initialization, \label{thisg}
  \item 1275 pairs, all of them treated with k-means but with different initializations, \label{thatg}
  \item 1275 pairs, all of them treated with single pass k-means but with different initializations.\label{g4} 
\end{enumerate}

\begin{figure}[h!]
   \centering
   \includegraphics[width=9cm]{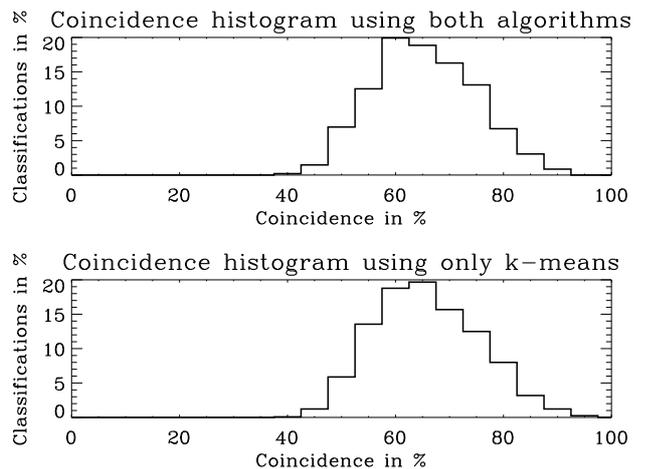}
      \caption{
Top: Histogram of coincidence for pairs of classifications, one 
inferred with k-means and the other with single pass k-means. The mean coincidence, 
around 70\,\%, is characteristic of the SDSS-DR7 galaxy spectrum
catalog \citep{2010ApJ...714..487S}.
Bottom:
Same as avobe, except that only classifications using
k-means are compared. It shows the intrinsic dispersion 
in possible classifications due to the random initialization 
of the algorithm.
              }
         \label{test}
   \end{figure}

The histograms with the values of coincidence for 
the groups \#~\ref{thisg} and \ref{thatg}
are represented in Fig. \ref{test}.
Both are very similar, including their means and standard deviations
($67.9\pm 9.4\,\%$ and $68.3\pm 9.5\,\%$, respectively). 
The lower histogram compares  classifications
derived using k-means alone, and so it quantifies the intrinsic scatter due 
to the random initialization. The upper histogram compares classifications from 
single pass k-means and k-means, so it includes the intrinsic scatter plus
the systematic differences that k-means and single pass k-means may have. 
Since the two distributions are so similar, 
we conclude that there are no systematic differences,
and the classifications inferred from  k-means and single pass k-means 
are equivalent.
The distributions corresponding to groups \#~\ref{g1} and \#~\ref{g4} are 
not shown, but they are very similar to those in Fig. \ref{test},
and from them it also follows that the classes inferred
from k-means  and single pass k-means are equivalent for 
practical applications.

The discussion above is purely qualitative. We have gone a step further to show 
that the histograms of coincidence corresponding to the four groups 
are statistically equivalent. The  Kolmogorov-Smirnov (KS) test allows 
determining the probability that two observed distributions are drawn 
from the same parent distribution \citep[e.g.,][]{Massey51}. 
Using the KS test, the probability that the histograms in Fig.~\ref{test} 
represent the same distribution is more than 99.9\%.  
Using all possible pairs of the histograms from the four groups,
the KS conclude that  the probability of being the same distribution
 is between 97\% and 100\%. Our claim that single pass k-means and
 k-means provide statistically equivalent classifications relies on this result.

%

%
%
\section{Conclusions}\label{conclusions}

The classification algorithm k-means has the potential to classify huge astronomical
databases, such as those to be expected with the advent of new instruments and 
catalogs (see Sect.~\ref{intro}).
We tested a variant of the original algorithm, called single pass k-means, which
unifies the two main steps of k-means 
(Sect.~\ref{algorithms}).
Single pass k-means turns out to be between 20\,\% and 40\,\% faster than k-means (Sect.~\ref{test1}),
and it provides statistically equivalent classifications (Sect.~\ref{equivalence}).

Saving 20\,\%\ to 40\,\%\ of the time may not look like a lot, 
however the actual gain
when using single pass k-means depends very much on the specific application. 
Keep in mind that k-means (and so single pass k-means) is a tool with the potential
of classifying gigantic datasets by bruteforce. The foreseeable applications may 
require  long execution times and, therefore a 40\,\%\ saving may actually 
represent days or weeks of work.

The tests were carried out using a particular catalog of galaxy spectra with 
limited data volumes (up to 20\,000 objects in 1637 dimensions). 
However, single pass k-means would outperform k-mean even 
for other larger datasets. That the computer time employed 
by the two alternative algorithms increases linearly with time
implies that the gain should be constant even for 
significantly larger datasets. Moreover, k-means is a workhorse 
proven to converge in many very different contexts. 
The datasets we use are not special, therefore the properties inferred for 
them can probably be extrapolated to many other datasets.

\bibliography{ms}

\begin{thebibliography}{18}
\expandafter\ifx\csname natexlab\endcsname\relax\def\natexlab#1{#1}\fi

\bibitem[{{Abazajian} {et~al.}(2009){Abazajian}, {Adelman-McCarthy},
  {Ag{\"u}eros}, {Allam}, {Allende Prieto}, {An}, {Anderson}, {Anderson},
  {Annis}, {Bahcall}, \& et~al.}]{2009ApJS..182..543A}
{Abazajian}, K.~N., {Adelman-McCarthy}, J.~K., {Ag{\"u}eros}, M.~A., {et~al.}
  2009, \apjs, 182, 543

\bibitem[{{Ascasibar} \& {S{\'a}nchez Almeida}(2011)}]{2011MNRAS.415.2417A}
{Ascasibar}, Y. \& {S{\'a}nchez Almeida}, J. 2011, \mnras, 415, 2417

\bibitem[{{Balazs} {et~al.}(1996){Balazs}, {Garibjanyan}, {Mirzoyan},
  {Hambaryan}, {Kun}, {Fronto}, \& {Kelemen}}]{1996A&A...311..145B}
{Balazs}, L.~G., {Garibjanyan}, A.~T., {Mirzoyan}, L.~V., {et~al.} 1996, \aap,
  311, 145

\bibitem[{Bishop(2006)}]{Bishop2006}
Bishop, C.~M. 2006, Pattern Recognition and Machine Learning (Springer), 424

\bibitem[{{Chattopadhyay} {et~al.}(2012){Chattopadhyay}, {Sharina}, {Davoust},
  {De}, \& {Chattopadhyay}}]{2012ApJ...750...91C}
{Chattopadhyay}, T., {Sharina}, M., {Davoust}, E., {De}, T., \&
  {Chattopadhyay}, A.~K. 2012, \apj, 750, 91

\bibitem[{{Fraix-Burnet} {et~al.}(2012){Fraix-Burnet}, {Chattopadhyay},
  {Chattopadhyay}, {Davoust}, \& {Thuillard}}]{2012A&A...545A..80F}
{Fraix-Burnet}, D., {Chattopadhyay}, T., {Chattopadhyay}, A.~K., {Davoust}, E.,
  \& {Thuillard}, M. 2012, \aap, 545, A80

\bibitem[{{Ivezic} {et~al.}(2008){Ivezic}, {Tyson}, {Acosta}, {Allsman},
  {Anderson}, {Andrew}, {Angel}, {Axelrod}, {Barr}, {Becker}, {Becla},
  {Beldica}, {Blandford}, {Bloom}, {Borne}, {Brandt}, {Brown}, {Bullock},
  {Burke}, {Chandrasekharan}, {Chesley}, {Claver}, {Connolly}, {Cook},
  {Cooray}, {Covey}, {Cribbs}, {Cutri}, {Daues}, {Delgado}, {Ferguson},
  {Gawiser}, {Geary}, {Gee}, {Geha}, {Gibson}, {Gilmore}, {Gressler}, {Hogan},
  {Huffer}, {Jacoby}, {Jain}, {Jernigan}, {Jones}, {Juric}, {Kahn}, {Kalirai},
  {Kantor}, {Kessler}, {Kirkby}, {Knox}, {Krabbendam}, {Krughoff}, {Kulkarni},
  {Lambert}, {Levine}, {Liang}, {Lim}, {Lupton}, {Marshall}, {Marshall}, {May},
  {Miller}, {Mills}, {Monet}, {Neill}, {Nordby}, {O'Connor}, {Oliver},
  {Olivier}, {Olsen}, {Owen}, {Peterson}, {Petry}, {Pierfederici},
  {Pietrowicz}, {Pike}, {Pinto}, {Plante}, {Radeka}, {Rasmussen}, {Ridgway},
  {Rosing}, {Saha}, {Schalk}, {Schindler}, {Schneider}, {Schumacher}, {Sebag},
  {Seppala}, {Shipsey}, {Silvestri}, {Smith}, {Smith}, {Strauss}, {Stubbs},
  {Sweeney}, {Szalay}, {Thaler}, {Vanden Berk}, {Walkowicz}, {Warner},
  {Willman}, {Wittman}, {Wolff}, {Wood-Vasey}, {Yoachim}, {Zhan}, \& {for the
  LSST Collaboration}}]{2008arXiv0805.2366I}
{Ivezic}, Z., {Tyson}, J.~A., {Acosta}, E., {et~al.} 2008, ArXiv e-prints

\bibitem[{Massey(1951)}]{Massey51}
Massey, F.~J. 1951, Journal of the American Statistical Association, 46, 68

\bibitem[{{Morales-Luis} {et~al.}(2011){Morales-Luis}, {S{\'a}nchez Almeida},
  {Aguerri}, \& {Mu{\~n}oz-Tu{\~n}{\'o}n}}]{2011ApJ...743...77M}
{Morales-Luis}, A.~B., {S{\'a}nchez Almeida}, J., {Aguerri}, J.~A.~L., \&
  {Mu{\~n}oz-Tu{\~n}{\'o}n}, C. 2011, \apj, 743, 77

\bibitem[{{Prusti}(2012)}]{2012AN....333..453P}
{Prusti}, T. 2012, Astronomische Nachrichten, 333, 453

\bibitem[{{Rebbapragada} {et~al.}(2009){Rebbapragada}, {Protopapas}, {Brodley},
  \& {Alcock}}]{Rebbapragada:2009:FAP:1507556.1507573}
{Rebbapragada}, U., {Protopapas}, P., {Brodley}, C.~E., \& {Alcock}, C. 2009,
  Mach. Learn., 74, 281

\bibitem[{{S{\'a}nchez Almeida} {et~al.}(2013){S{\'a}nchez Almeida}, {Aguerri},
  \& {Mu{\~n}oz-Tu{\~n}{\'o}n}}]{2013RMxAC..42..111S}
{S{\'a}nchez Almeida}, J., {Aguerri}, J.~A.~L., \& {Mu{\~n}oz-Tu{\~n}{\'o}n},
  C. 2013, in Revista Mexicana de Astronomia y Astrofisica Conference Series,
  Vol.~42, Revista Mexicana de Astronomia y Astrofisica Conference Series,
  111--111

\bibitem[{{S{\'a}nchez Almeida} {et~al.}(2010){S{\'a}nchez Almeida}, {Aguerri},
  {Mu{\~n}oz-Tu{\~n}{\'o}n}, \& {de Vicente}}]{2010ApJ...714..487S}
{S{\'a}nchez Almeida}, J., {Aguerri}, J.~A.~L., {Mu{\~n}oz-Tu{\~n}{\'o}n}, C.,
  \& {de Vicente}, A. 2010, \apj, 714, 487

\bibitem[{{S{\'a}nchez Almeida} {et~al.}(2009){S{\'a}nchez Almeida}, {Aguerri},
  {Mu{\~n}oz-Tu{\~n}{\'o}n}, \& {Vazdekis}}]{2009ApJ...698.1497S}
{S{\'a}nchez Almeida}, J., {Aguerri}, J.~A.~L., {Mu{\~n}oz-Tu{\~n}{\'o}n}, C.,
  \& {Vazdekis}, A. 2009, \apj, 698, 1497

\bibitem[{{S{\'a}nchez Almeida} \& {Allende
  Prieto}(2013)}]{2013ApJ...763...50S}
{S{\'a}nchez Almeida}, J. \& {Allende Prieto}, C. 2013, \apj, 763, 50

\bibitem[{{S{\'a}nchez Almeida} \& {Lites}(2000)}]{2000ApJ...532.1215S}
{S{\'a}nchez Almeida}, J. \& {Lites}, B.~W. 2000, \apj, 532, 1215

\bibitem[{{S{\'a}nchez Almeida} {et~al.}(2012){S{\'a}nchez Almeida},
  {Terlevich}, {Terlevich}, {Cid Fernandes}, \&
  {Morales-Luis}}]{2012ApJ...756..163S}
{S{\'a}nchez Almeida}, J., {Terlevich}, R., {Terlevich}, E., {Cid Fernandes},
  R., \& {Morales-Luis}, A.~B. 2012, \apj, 756, 163

\bibitem[{{Simpson} {et~al.}(2012){Simpson}, {Cottrell}, \&
  {Worley}}]{2012MNRAS.427.1153S}
{Simpson}, J.~D., {Cottrell}, P.~L., \& {Worley}, C.~C. 2012, \mnras, 427, 1153

\end{thebibliography}
\bibliographystyle{aa}%

\end{document}